\documentclass[twocolumn]{aastex62}
\usepackage{amsmath}

\received{\today}
\revised{\today}
\accepted{-----}
\submitjournal{----}

\shorttitle{Halo Rotation}
\shortauthors{Tian et al.}
\begin{document}
\title{Rotating halo traced by the K giant stars from LAMOST and Gaia}

\correspondingauthor{Chao Liu}
\email{liuchao@nao.cas.cn}

\author[0000-0003-3347-7596]{Hao Tian}
\affil{Key Laboratory of Optical Astronomy, \\
National Astronomical Observatories, 
Chinese Academy of Sciences, \\Datun Road 20A,
Beijing 100012, PR China;}
\collaboration{(LAMOST Fellow)}

\author[0000-0002-1802-6917]{Chao Liu}
\affiliation{Key Laboratory of Optical Astronomy, \\
National Astronomical Observatories, Chinese Academy of Sciences, \\Datun Road 20A,
Beijing 100012, PR China;}

\author{Yan Xu}
\affiliation{Key Laboratory of Optical Astronomy, \\
National Astronomical Observatories, Chinese Academy of Sciences, \\Datun Road 20A,
Beijing 100012, PR China;}

\author{Xiangxiang Xue}
\affiliation{Key Laboratory of Optical Astronomy, \\
National Astronomical Observatories, Chinese Academy of Sciences, \\Datun Road 20A,
Beijing 100012, PR China;}

\begin{abstract}
With the help of Gaia DR2, we are able to obtain the full 6-D phase space 
information for stars from LAMOST DR5. With high precision of position, velocity, 
and metallicity, the rotation of the local stellar halo is presented using the 
K giant stars with [Fe/H]$<-1$\,dex  within 4 kpc from the Sun. By fitting the rotational velocity 
distribution with three Gaussian components,  stellar halo,  disk, and a 
counter-rotating hot population, we find that the local halo progradely rotates with 
$V_T=+27^{+4}_{-5}$\,km\ s$^{-1}$ providing the local standard of rest velocity of 
$V_{LSR}=232$\,km\ s$^{-1}$. Meanwhile, we obtain the dispersion of rotational 
velocity is $\sigma_{T}=72^{+4}_{-4}$\,km s$^{-1}$.
Although the rotational velocity strongly depends on the choice of $V_{LSR}$, 
the trend of prograde rotation is substantial even when $V_{LSR}$ is set at as 
low as 220\,km\ s$^{-1}$. Moreover, we derive the rotation for subsamples with different 
metallicities and find that the rotational velocity is 
essentially not correlated with [Fe/H]. 
This may hint a secular evolution origin of the prograde rotation.
It shows that the metallicity of the progradely rotating
 halo is peaked within -1.9$<$[Fe/H]$<$-1.6 
without considering the selection effect.
We also find a small fraction of counter-rotating stars 
with larger dispersion and lower metallicity.
Finally, the disk component rotates with $V_T=+182^{+6}_{-6}$\,km\ s$^{-1}$ and 
$\sigma_T=45^{+3}_{-3}$\,km\ s$^{-1}$, which is quite consistent 
with the metal-weak thick disk population.

\end{abstract}


\section{Introduction} \label{sec:intro}

It has been widely accepted that the Milky Way, at least the halo, was formed
through hierarchically merging smaller stellar systems,
 such as globular clusters
and dwarf galaxies \citep{2006ApJ...642L.137B}. 
With a long time evolving, the stellar members of those merged
systems will be continuously stripped and
 then form stellar streams \citep{1999ApJ...512L.109J}. Such 
phenomena have been widely searched with photometric sky surveys, e.g.
SDSS \citep{2000AJ....120.1579Y} and Pan-STARRS \citep{2016MNRAS.463.1759B}. 
Since the first stellar stream, Sagittarius stream, discovered by 
\citet{1994Natur.370..194I}, many streams have been discovered with matched filter 
method 
\citep{2002AJ....124..349R, 2016ASSL..420...87G, 
2006ApJ...641L..37G, 2006ApJ...643L..17G, 2006AAS...208.4803G, 2009ApJ...693.1118G} 
and other methods \citep{2018MNRAS.474.4112M}.
But the streams are not the final fate of those merged clusters or dwarf galaxies,
a longer evolving time will make the streams more and more relaxed and finally become 
part of the smoothing halo, i.e. it is not visible in the geometric 
space \citep{1999Natur.402...53H}, even blurred in the 
phase space \citep{2000MNRAS.319..657H, 2015ApJ...801...98S}. 

Assuming those merged stellar systems fell in with random orbits 
\citep{2015ApJ...801...98S},  the contribution of $z-$axis angular 
momentum $L_z$ overall should be around 0. The net angular mometum is directly
related to the merging history,  not only including the minor mergers of the globular 
cluster and dwarf galaxies, such as the Sagittarius Stream, but also including the 
major merger events, such as the recently discovered 
Gaia-Enceladus\citep{2018Natur.563...85H}, also named 
\emph{Gaia-Sausage}\citep{2018MNRAS.478..611B}, which contibutes
a larger fraction of retrogradely rotating stars.
Another possibility is that 
the merged halo was driven to slowly rotate by the non-axisymmetric structure
in the disk (e.g. bar) in a long time scale \citep{2013MNRAS.429.1949A}.

 The studies on dynamics of the halo have been done since 
1980s \citep{1980MNRAS.193..295F} who used 66 clusters to constrained the rotation
 velocity 60$\pm$26 km s$^{-1}$.
As the velocity ellipsoid is directly related to the 
mass distribution of the halo \citep{2008gady.book.....B},
more work focused on the distribution of the 
dispersions \citep{1997ApJ...481..775S, 2004AJ....127..914S,
2009MNRAS.399.1223S, 2013ApJ...763L..17H, 2013MNRAS.430.2973K, 2018ApJ...853..196L}.

Because of the difficulty of observation of proper motions, especially for distant stars.
\citet{1990AJ....100.1191M} studied the kinematics
of the halo and disk using the G and K giant stars within 4 kpc from the Sun, and claimed
a prograde rotation of $V_T=25\pm15$ km s$^{-1}$. They also showed that the metallicity 
of the thick disk stars can be as low as -1.6.
\citet{2009MNRAS.399.1223S} studied the 
kinematics of the  local halo  using $\sim$1700 solar neighbourhood subdwarfs
and found that the stellar halo exhibits no rotation  and the
rotational velocity $V_{\phi}$ dispersion $82\pm2$ km s$^{-1}$ , what should be noticed 
is that the typical velocity errors are large, about 30 to 50 km s$^{-1}$.

 More recently, the progress of the observations and data analysis techniques,
high precision proper motions have been obtained through combining catalogues obtained at different 
epochs cross a long time baseline \citep{2017ApJS..232....4T, 2018MNRAS.478..611B}.
 \citet{2017MNRAS.470.1259D} used the 3D velocity
calculated from the proper motion of different tracers, including RR Lyrae,
Blue Horizontal Branch (BHB) and K giant stars, in a distant volume 
from \emph{Gaia} Data Release (\emph{hereafter GDR}) 1 and SDSS
dataset to study the halo rotation. A prograde rotation was found 
with $V_T=14\pm2\pm10$ km s$^{-1}$. What is more, the metal richer stars are
showing a stronger prograde rotation which is explained as  a small fraction of
halo stars formed \emph{in situ}.
\citet{2017MNRAS.470.2959K} also found a similar trend with samples of 
Main Sequence Turnoff and K giant stars only with radial velocity avalible. 
What is different is that the metal poorer K giant stars show a retrograde rotation. 

 Since \citet{2007Natur.450.1020C} discovered the inner and outer halo and claimed that 
the inner halo is rotating progradely, in contrast the outer components is rotating
retrogradely. But the systematic bias in distance estimates led to a
 debate \citep{2011MNRAS.415.3807S, 2012ApJ...746...34B}.
This will be highly improved by the Gaia mission \citep{2001A&A...369..339P}. 

GDR2 \citep{2018arXiv180409365G, 2018arXiv180409378G, 
2018arXiv180409379G, 2018arXiv180409380G,
2018arXiv180409381G, 2018arXiv180409382G} contains 
$\sim$1.3 billion stars high 
accuracy proper motions and parallaxes, which have brought a 
series of amazing results about the halo,
 such as the velocity determination of globular clusters and 
rotation curve of the Large Magellanic Cloud \citep{2018arXiv180409381G}. 
Besides, \citet{2018arXiv180409381G} also derived a 
lower limit  total mass for the Milky Way.

Guoshoujing Telescope (the Large Sky Area Multi-Object Fiber Spectroscopic Telescope, 
hereafter LAMOST) is a 4-meter, quasi-meridian, reflecting Schmidt telescope
which makes it an efficient spectroscopic telescope \citep{2012RAA....12.1197C,
2012RAA....12..735D, 2012RAA....12..723Z}.
\cite{2017RAA....17...96L} and \cite{2018MNRAS.473.1244X} used metal poor 
K giants from LAMOST to trace the density profile of the halo and found an inner oblate,
outer nearly spherical profile.
Recently, LAMOST has released its 5th data (DR5). 
As low resolution of $R=1800$, the uncertainties of $T_{\rm eff}$ and [Fe/H] 
are typical $\sim100$\,K and $\sim0.1$\,dex, respectively, 
for the stellar spectra with signal to noise ratio larger than $30$.

The paper studies the rotation of the local stellar halo using the conbination of the 
GDR2 and LAMOST DR5 K-giant stars and is constructed as follow.
 In Section \ref{sect:data}, the sample stars are selected.
 The model of the distibution of rotational velociy is developed in 
 Section~\ref{sect:model}. The results are discussed in Section \ref{sect:result}.
The conclusions are listed in Section \ref{sect:conclusion}.
\section{Data Sample}
\label{sect:data}

\begin{figure*}
\centering
 \includegraphics[width=0.3\textwidth]{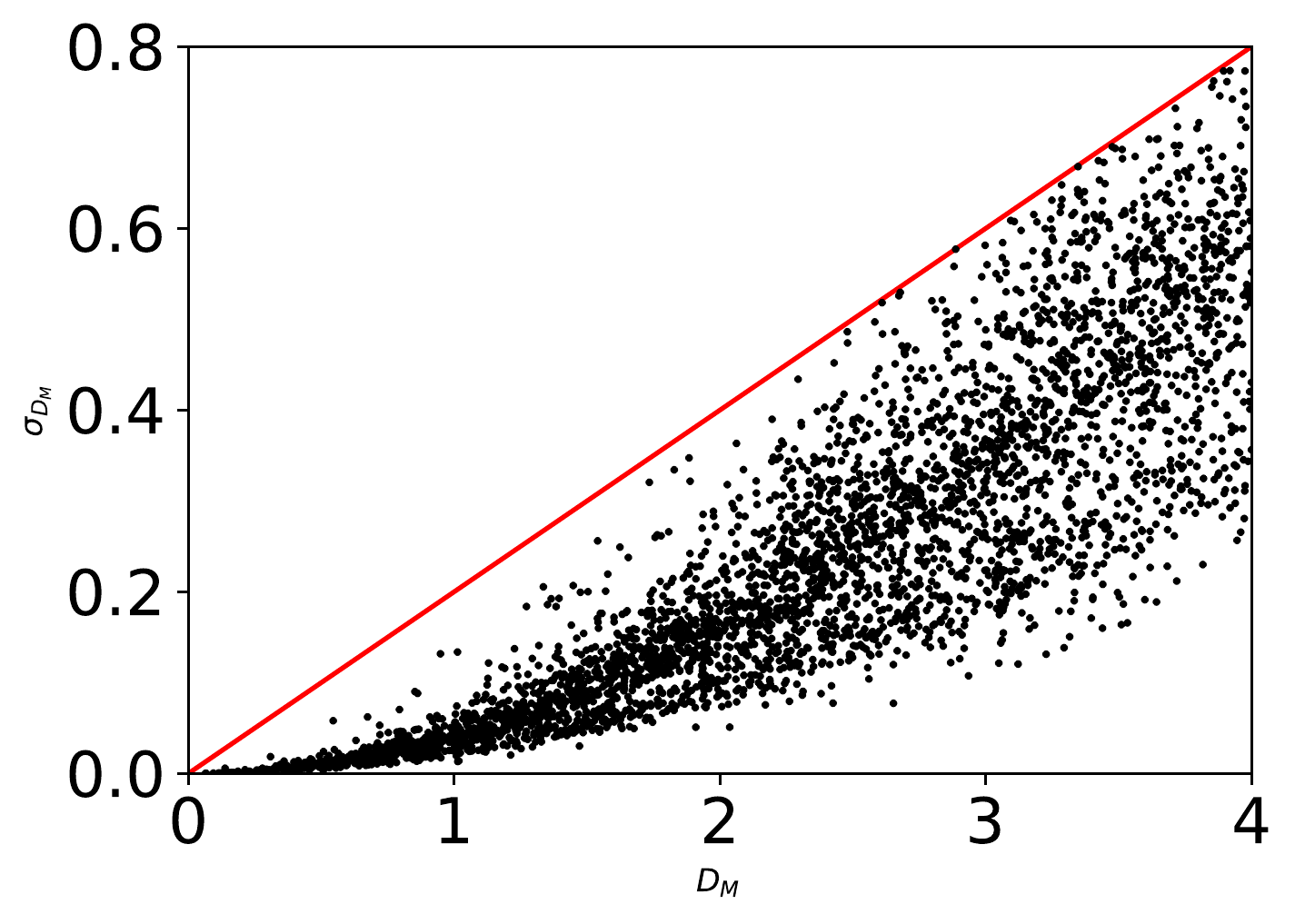}
 \includegraphics[width=0.3\textwidth]{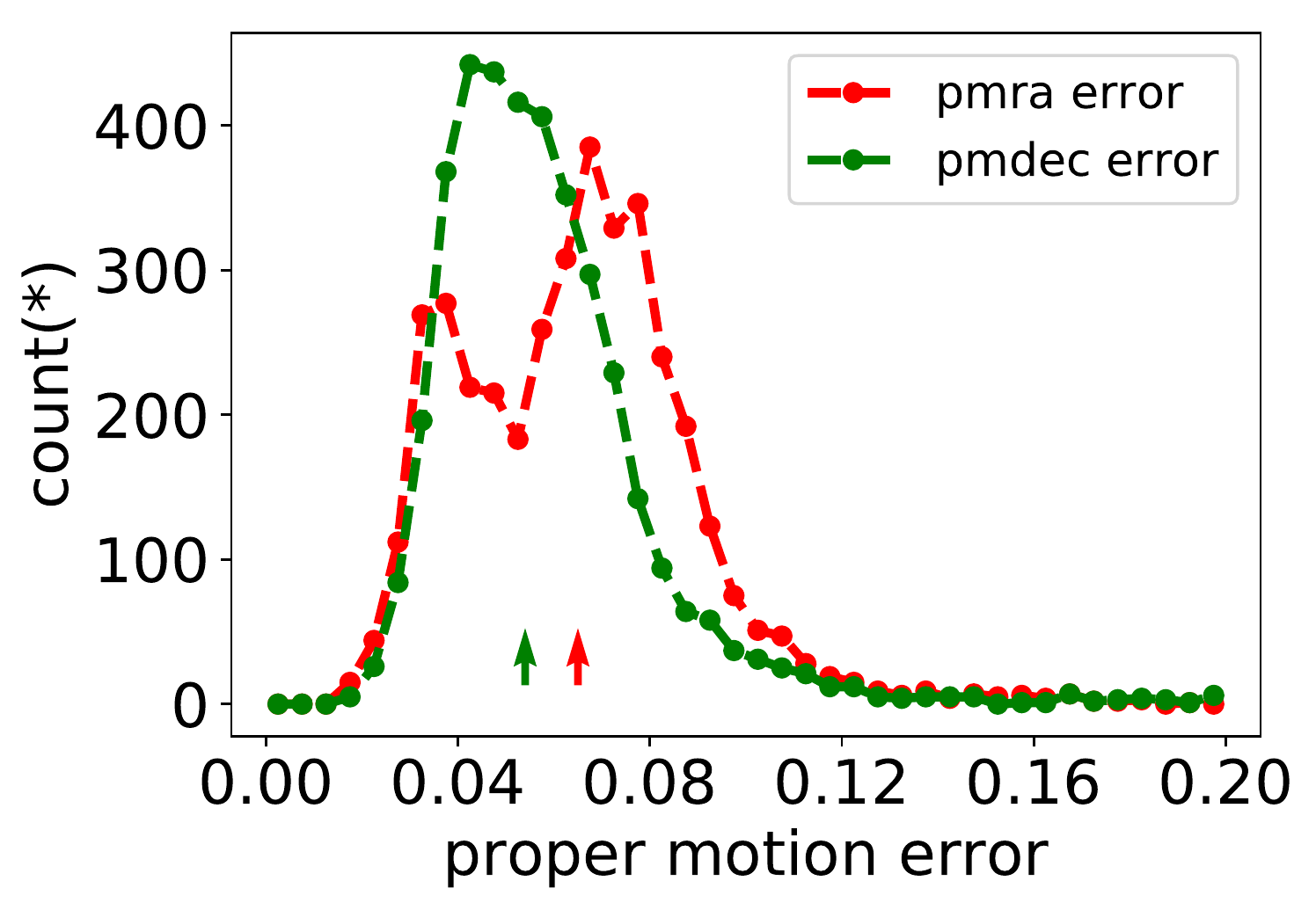}
 \includegraphics[width=0.3\textwidth]{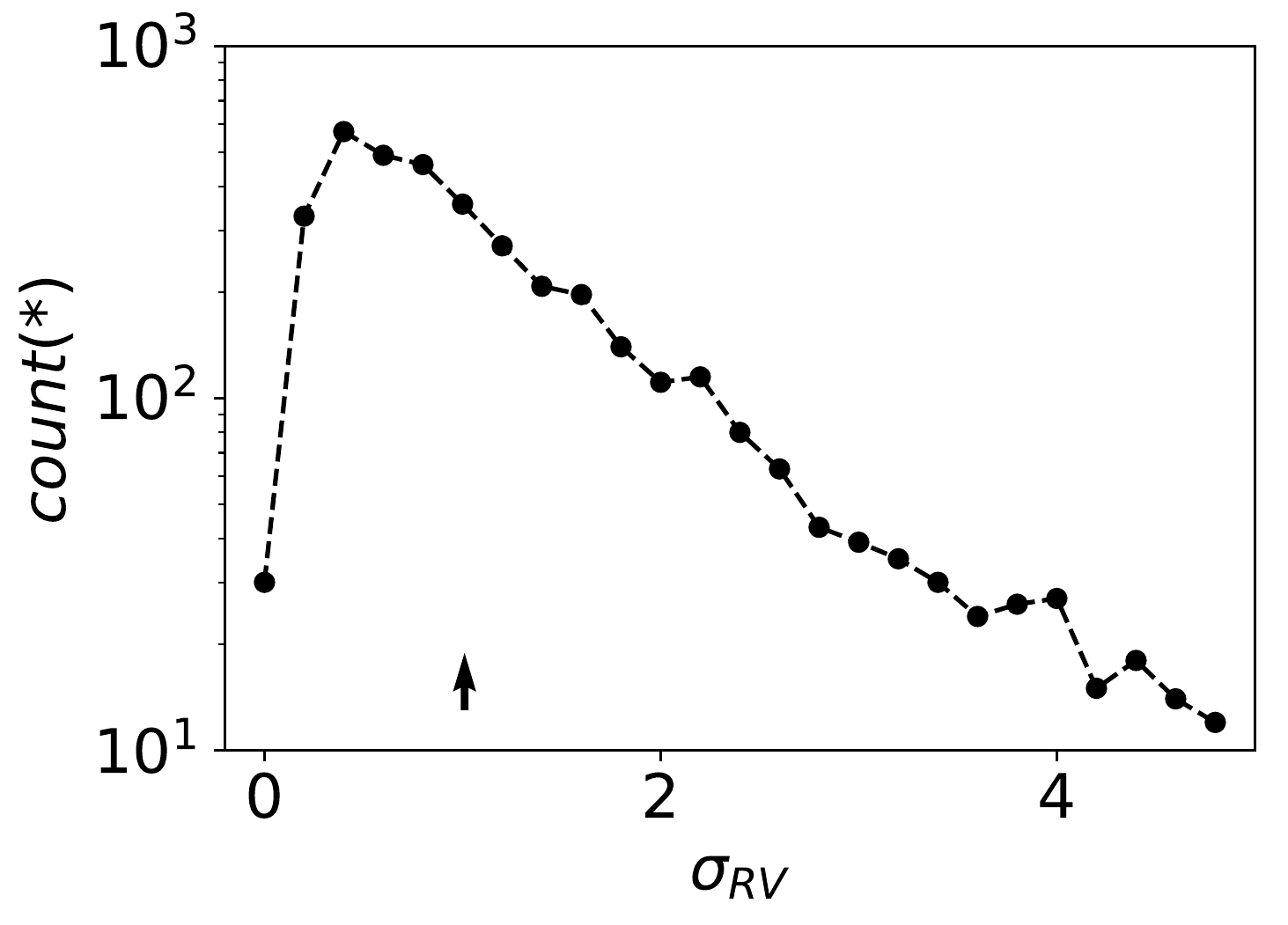}
 \caption{Distributions of the errors of proper motions, 
 radial velocities and cylindrical velocities
 in the left, middle and right panels respectively. 
 The red line in the left panel indicates the limit of the distance error 20\%.
 The arrows indicate the median errors for each 
 quantities. }
 \label{fig:Errors}
 \end{figure*}

In this work, we use K giant stars selected from LAMOST DR5 based on the criteria suggested by 
\citet{2014ApJ...790..110L} to study the  rotation of the nearby halo.
 After cross-matching with GDR2, we obtained the 
 parallax, proper motions, radial velocities and metallicities for the tracer 
 stars. 
A robust metallicity cut with [Fe/H]$<-1$ is used to remove the majority of disk stars, 
which leaves  6660 K giant stars.  In this case most of the thin disk stars and many thick disk stars will be removed.
As we will discuss in the following sections, the left disk component will be considered during the analysis.
With the distances from
 \cite{2018arXiv180410121B}, the samples within 4\,kpc and relative distance error
  $D_{error}/D<0.2$ are further selected so that the distance estimates are reliable,
  and  3827 stars are left  in the sample. 
  Comparing the radial velocities between GDR2 and LAMOST, we find an offset 
 $RV_{Gaia}-RV_{LAMOST}$ of 5.38 km s$^{-1}$,
  which was also mentioned by \citet{2015ApJ...809..145T} and \citet{2017MNRAS.472.3979S}.
   To avoid this systematics, we adopt the radial velocity provided by GDR2. 
 Figure~\ref{fig:Errors} shows the error distributions of distance, proper motions and radial velocity
in the left, middle and right panels respectively. We can find that the errors of distance increase 
with distances. From the middle panel, the median
 errors of proper motion in $RA$ and $DEC$ are 
 0.065 and 0.054 mas yr$^{-1}$, respectively, which are labelled with red and green arrows. The
 median error of radial velocity is 1.0 km s$^{-1}$ labelled with arrow in the right panel. 

We use \emph{galpy} \citep{2015ApJS..216...29B}
to calculate the 3D locations in heliocentric cartesian coordinates, $X$, $Y$, $Z$, 
and velocity components in galactocentric cylindrical coordinates, radial component
 $V_{R}$, rotational (azimuthal) component $V_{T}$, and vertical component $V_{Z}$.
By default, we adopt the velocity of the local standard of rest (LSR) with
 respect to the Galactic center as 
$V_{LSR}=232$ km s$^{-1}$ \citep{2017MNRAS.465...76M}
 and the solar motion with respect to the LSR as
$(U_{\odot},V_{\odot},W_{\odot})=$(11.1, 12.24, 7.25) km s$^{-1}$ 
\citep{2010MNRAS.403.1829S}.

\begin{figure}
\centering
\begin{minipage}{9cm}
 \centering
 \includegraphics[width=0.45\textwidth]{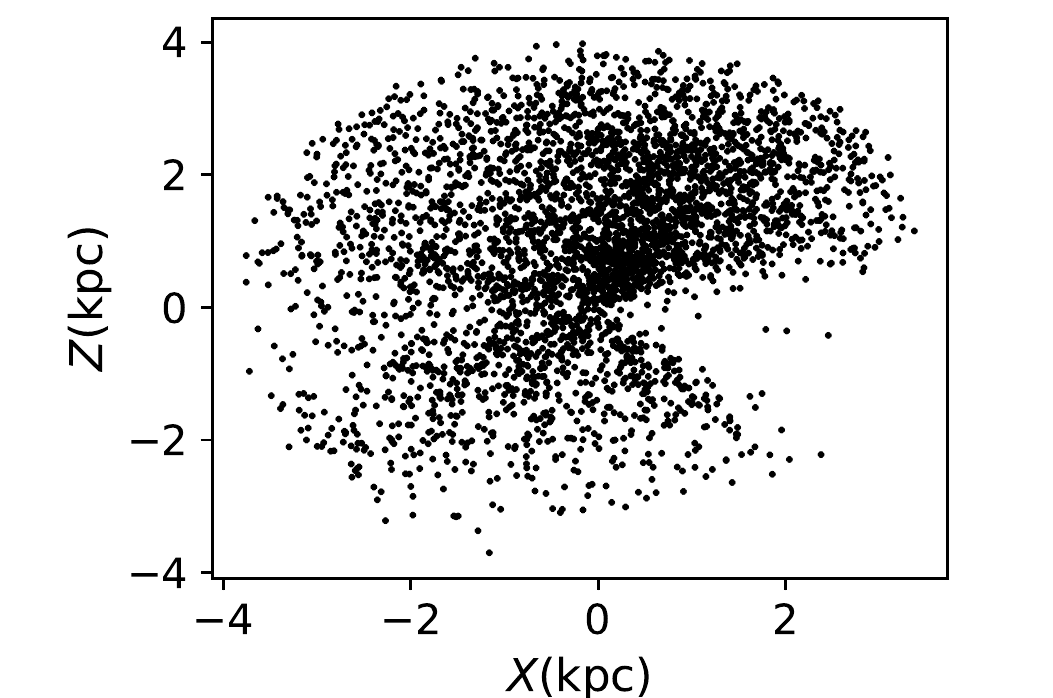}
 \includegraphics[width=0.45\textwidth]{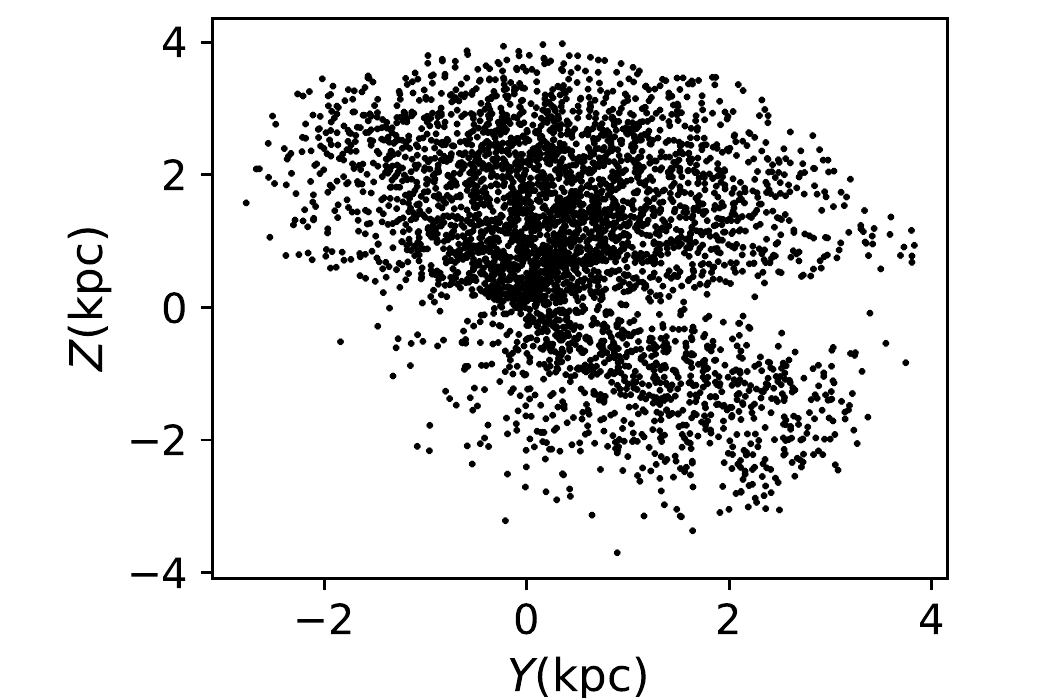}
 \includegraphics[width=0.45\textwidth]{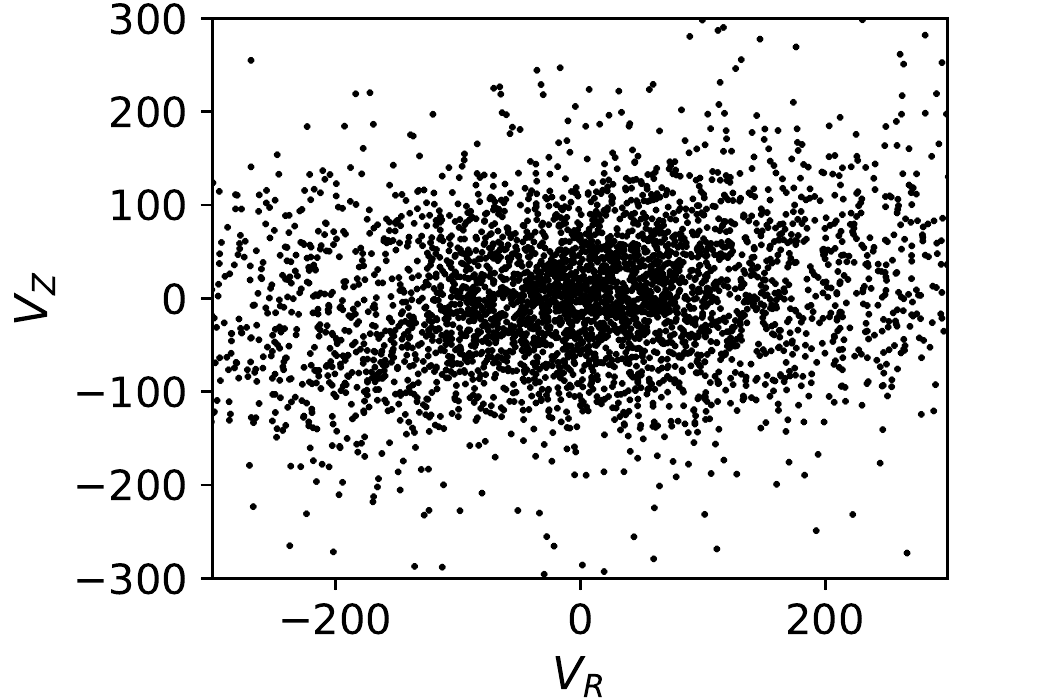}
 \includegraphics[width=0.45\textwidth]{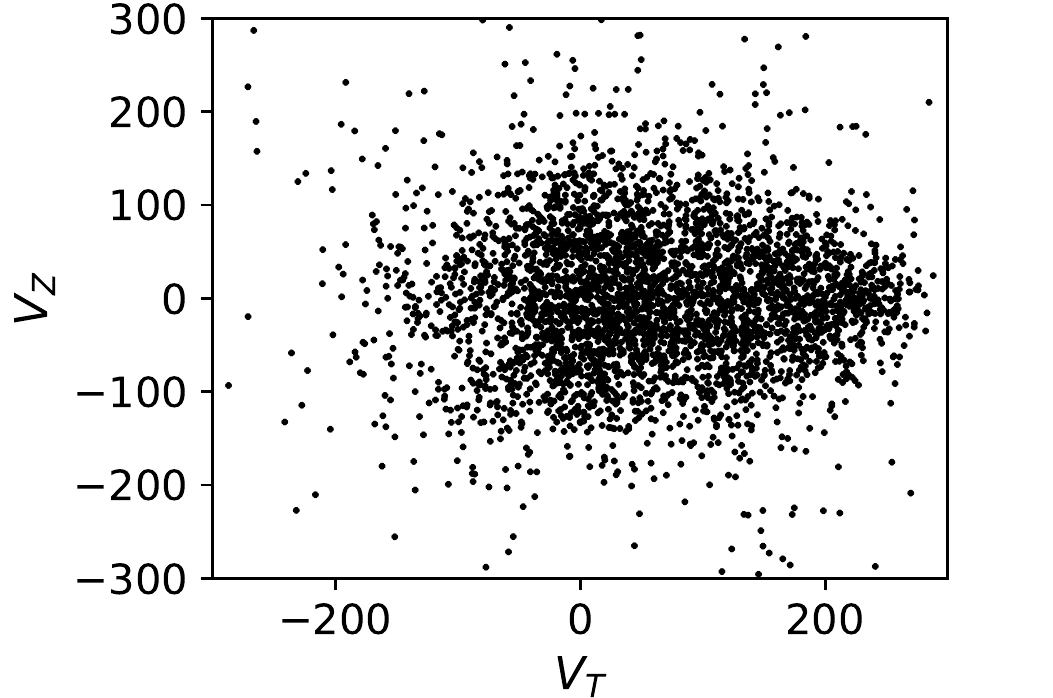}
 \end{minipage}
 \caption{Distributions of the sample in geometric space (top panels) and 
 velocity space (bottom panels).}
 \label{fig:pos_vel}
 \end{figure}
 
Figure~\ref{fig:pos_vel} shows the distributions of the selected K giant stars in 
geometric (top panels) and velocity (bottom panels) spaces. 
The samples do not show disk-like
 features, implying that most of the disk stars are removed by the cut in [Fe/H].
The top panels show that 
the stars are rarely sampled at $X>0$, $Y>0$, and $Z\sim0$, which is caused
by the observational limit of LAMOST survey. This would not affect the detection of the rotation
of the halo.
  The bottom-left panel shows that the distribution in $V_R$ versus $V_Z$ space is
 quite smooth. And the bottom-right panel displays an excess with $V_T$ at around 200 km s$^{-1}$,
 which may be the slight contamination of the thick disk.

 \begin{figure*}
 \centering
 \includegraphics[width=0.3\textwidth]{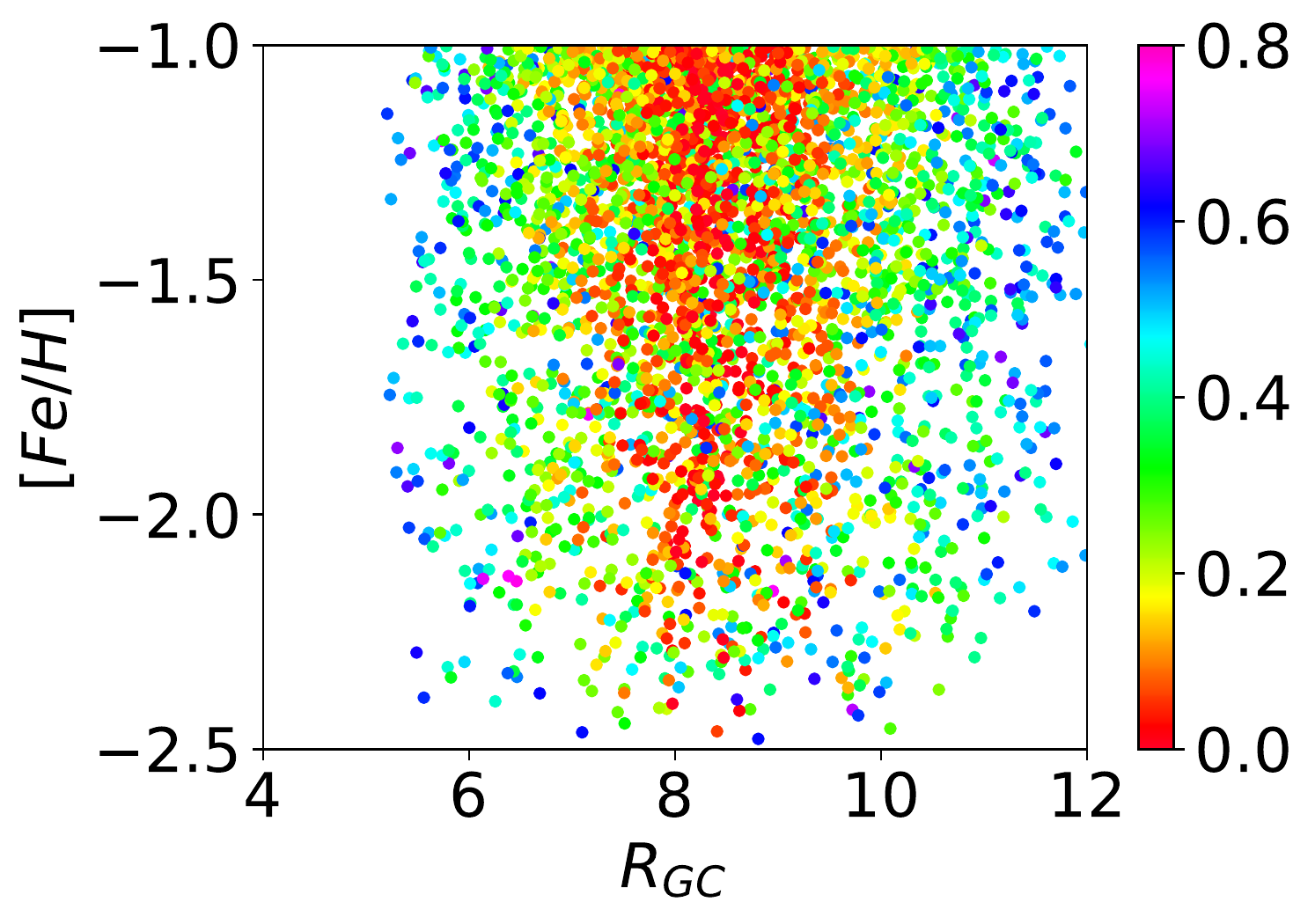}
 \includegraphics[width=0.3\textwidth]{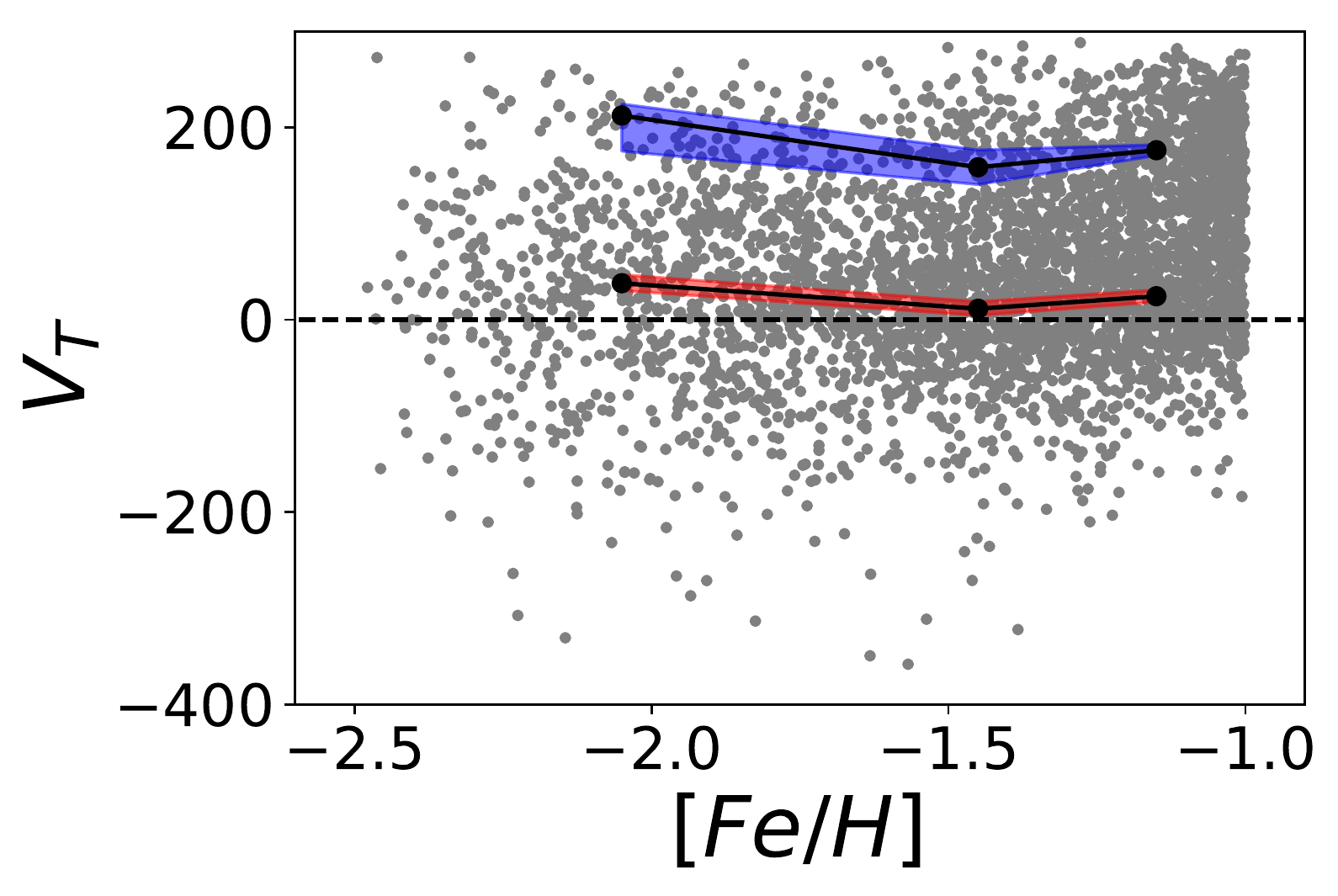}
 \includegraphics[width=0.3\textwidth]{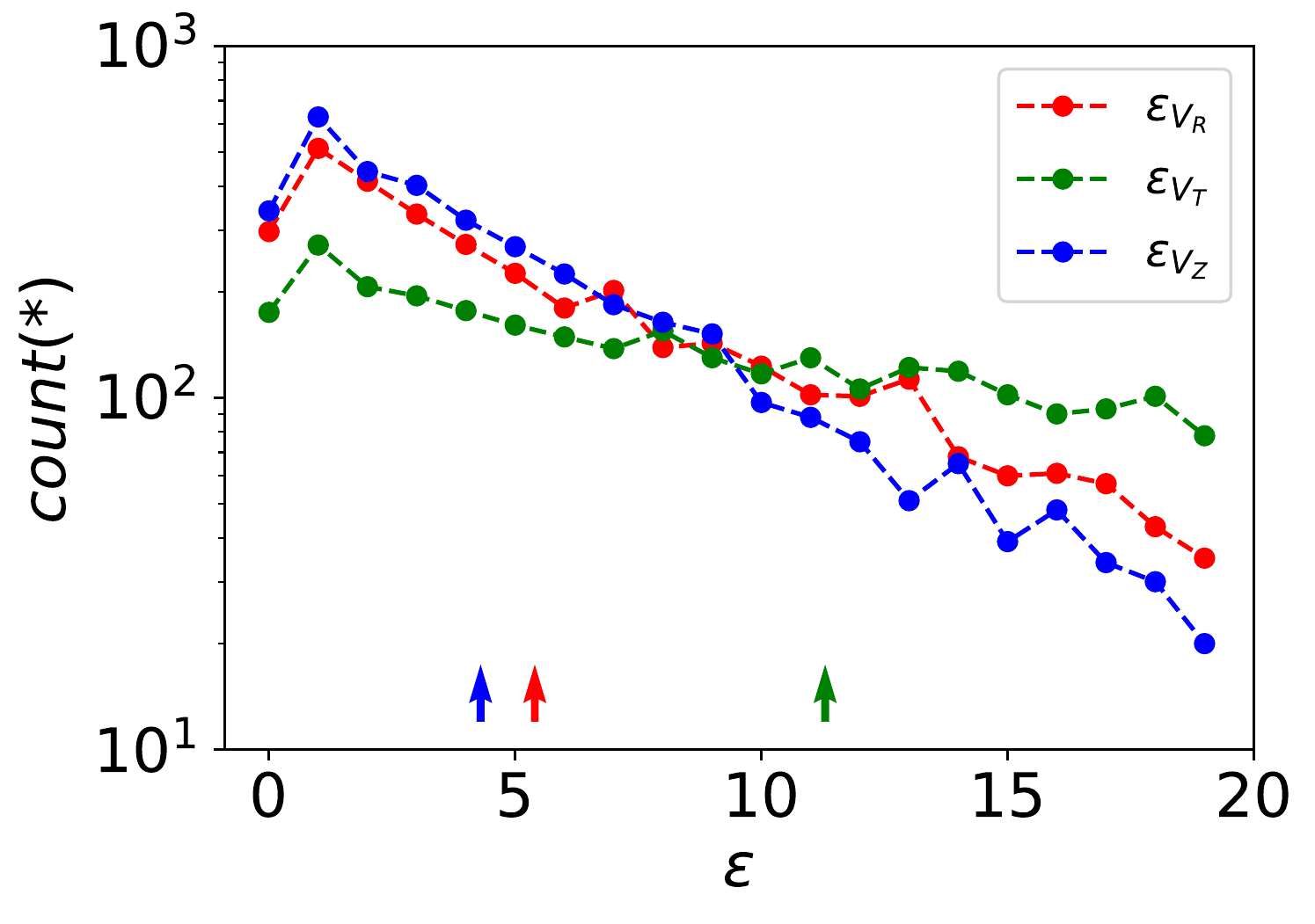}
 \caption{Left panel: The distribution of stars in $R_{GC}-[Fe/H]$ plane with 
 color-coded  distance error. Middle panel: The rotational velocity distribution 
 of the K giant stars is shown as
 the gray dots. The black solid lines in the blue and red shaded regions
 indicate the rotational velocities of the disk and halo components, respectively.
 The shadow colors from heavy to light represent the uncertainties from 1$\sigma$ 
 to 3$\sigma$. Right panel: The distribution of the errors of three velocity 
 components. The red, green and blue arrows indicate the median errors for 
 $V_R$, $V_T$ and $V_Z$, respectively.}
 \label{fig:Comparison}
 \end{figure*}

  In Figure~\ref{fig:Comparison}, noobvious correlation between $[Fe/H]$ and $R_{GC}$
 is found. The colors of the dots indicate the distance errors, which also show no significant 
 correlation with $[Fe/H]$. This implies that the samples do
  not experience any selection bias in metallicity.

 In the middle panel, the distribution 
 of the rotation velocity $V_T$ versus metallicity $[Fe/H]$ is shown with the gray dots. 
 Clearly we find that 
 more stars are progradely rotating in the metal richer side, which is suggested to be 
 the disk component. 
 The black lines in the blue and
 red shades are the rotation velocity for the halo and disk components respectively, which
 will be discussed in the later sections.

 In the right panel, the errors of the velocities are shown, which are calculated
 following the calculation of \citet{1987AJ.....93..864J}. The median errors of $V_R$, $V_T$ 
 and $V_Z$ are 5.4, 11.3 and 4.3 km s$^{-1}$ as represented by the red, green and blue arrows,
 respectively.

\section{Model of the distribution of $V_T$}
\label{sect:model}
\begin{figure*}
 \centering
 \begin{minipage}{18cm}
 \centering
 \includegraphics[width=0.45\textwidth]{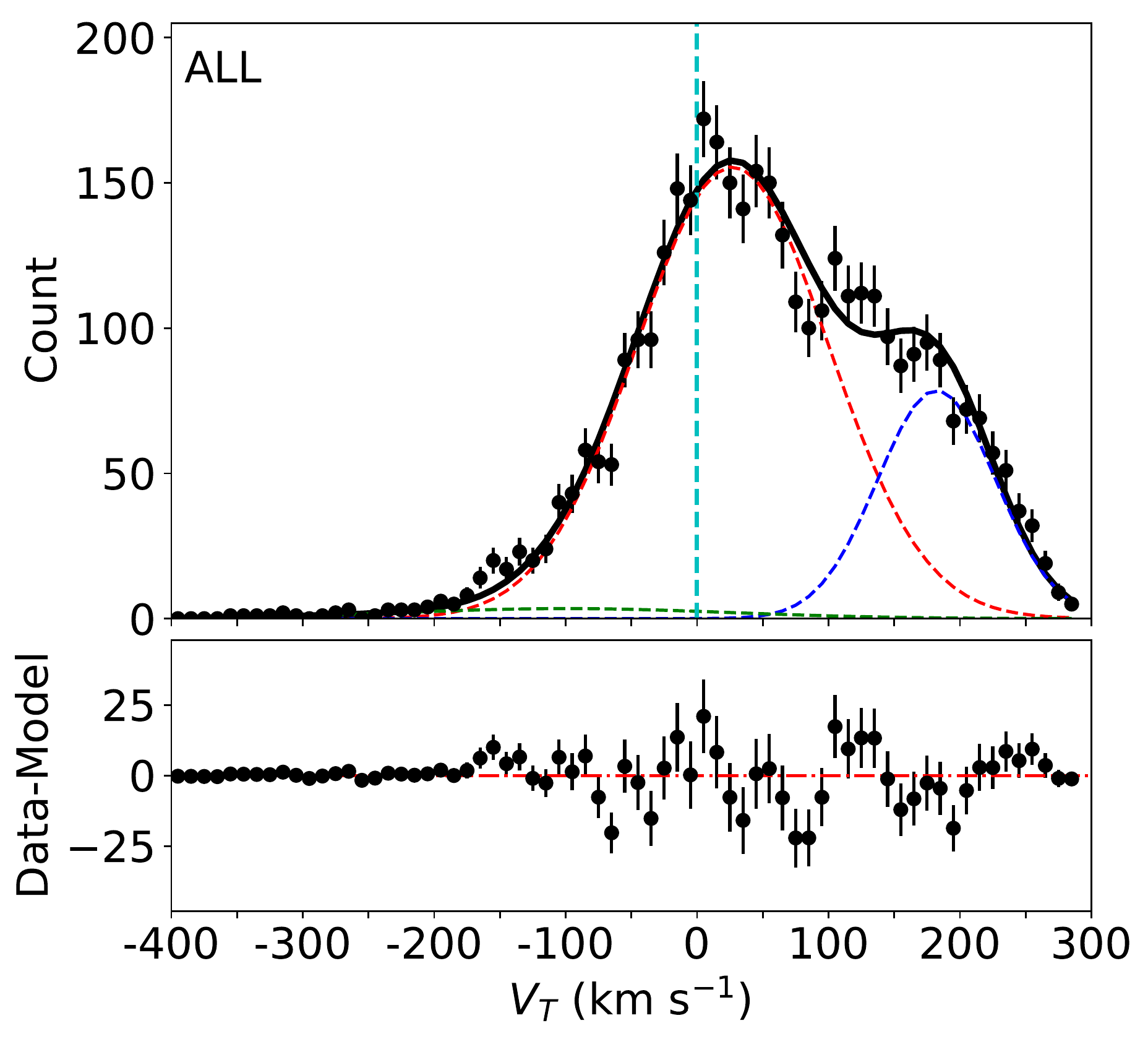}
 \includegraphics[width=0.45\textwidth]{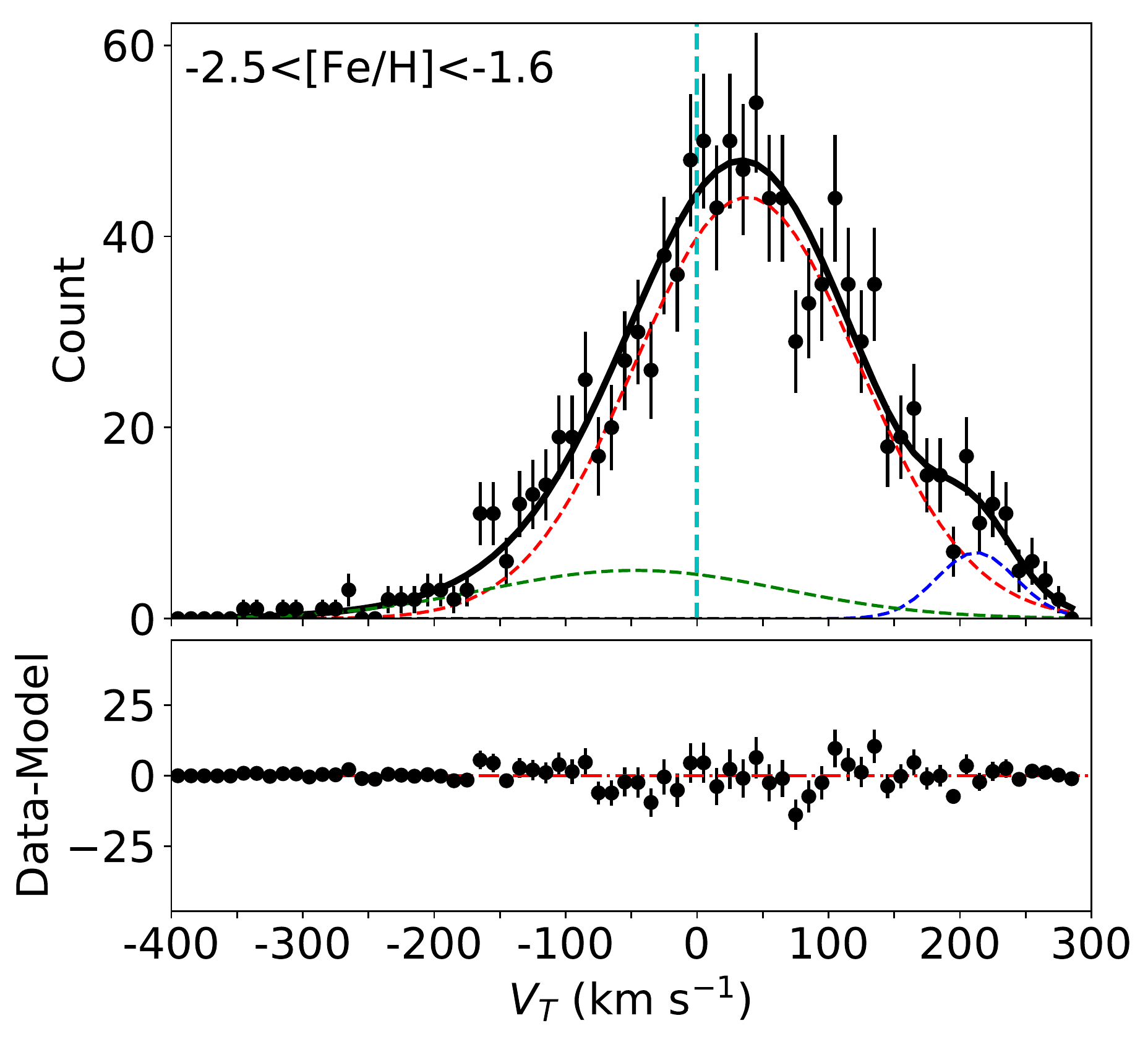}
  \end{minipage}
   \begin{minipage}{18cm}
   \centering
 \includegraphics[width=0.45\textwidth]{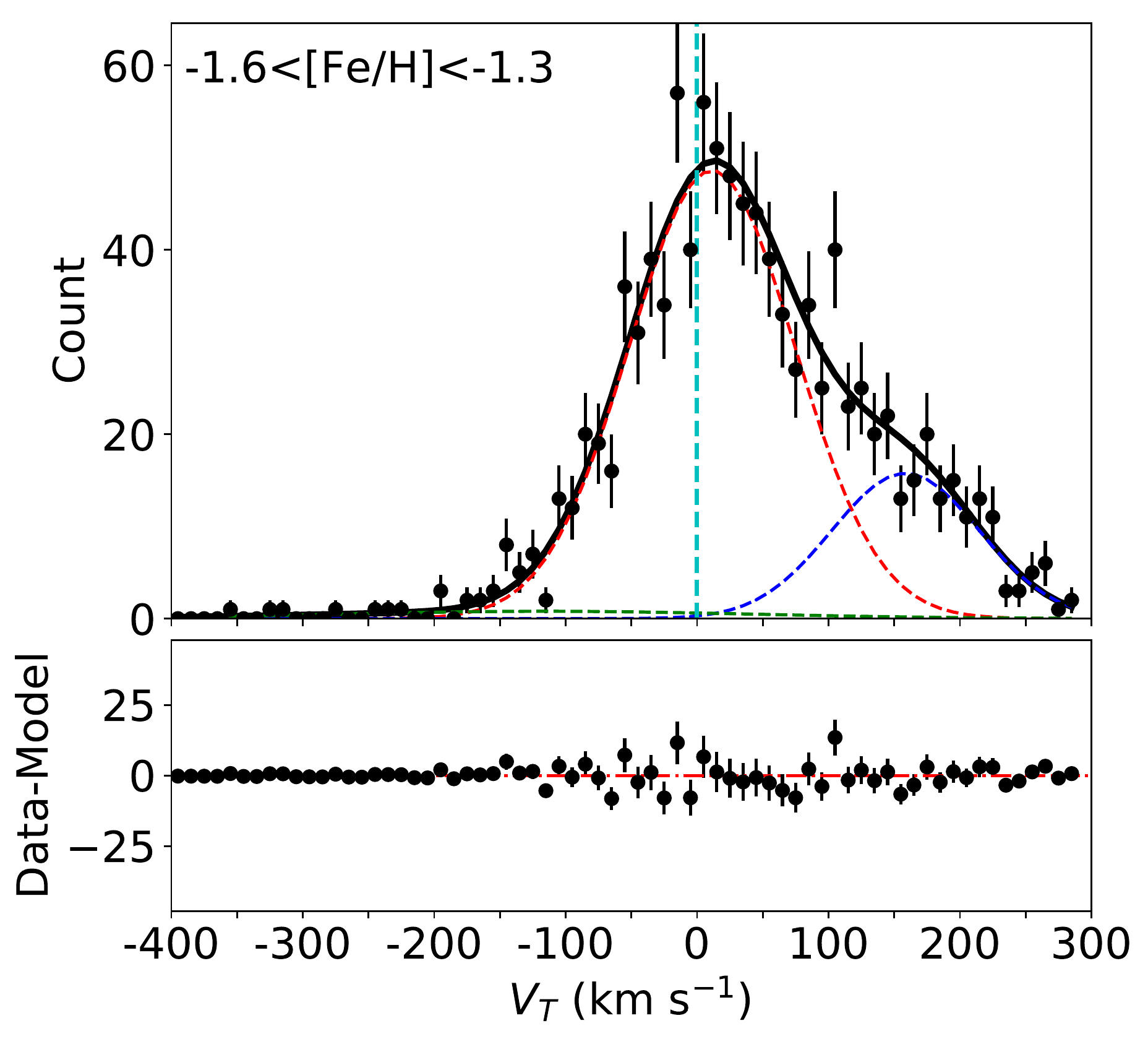}
 \includegraphics[width=0.45\textwidth]{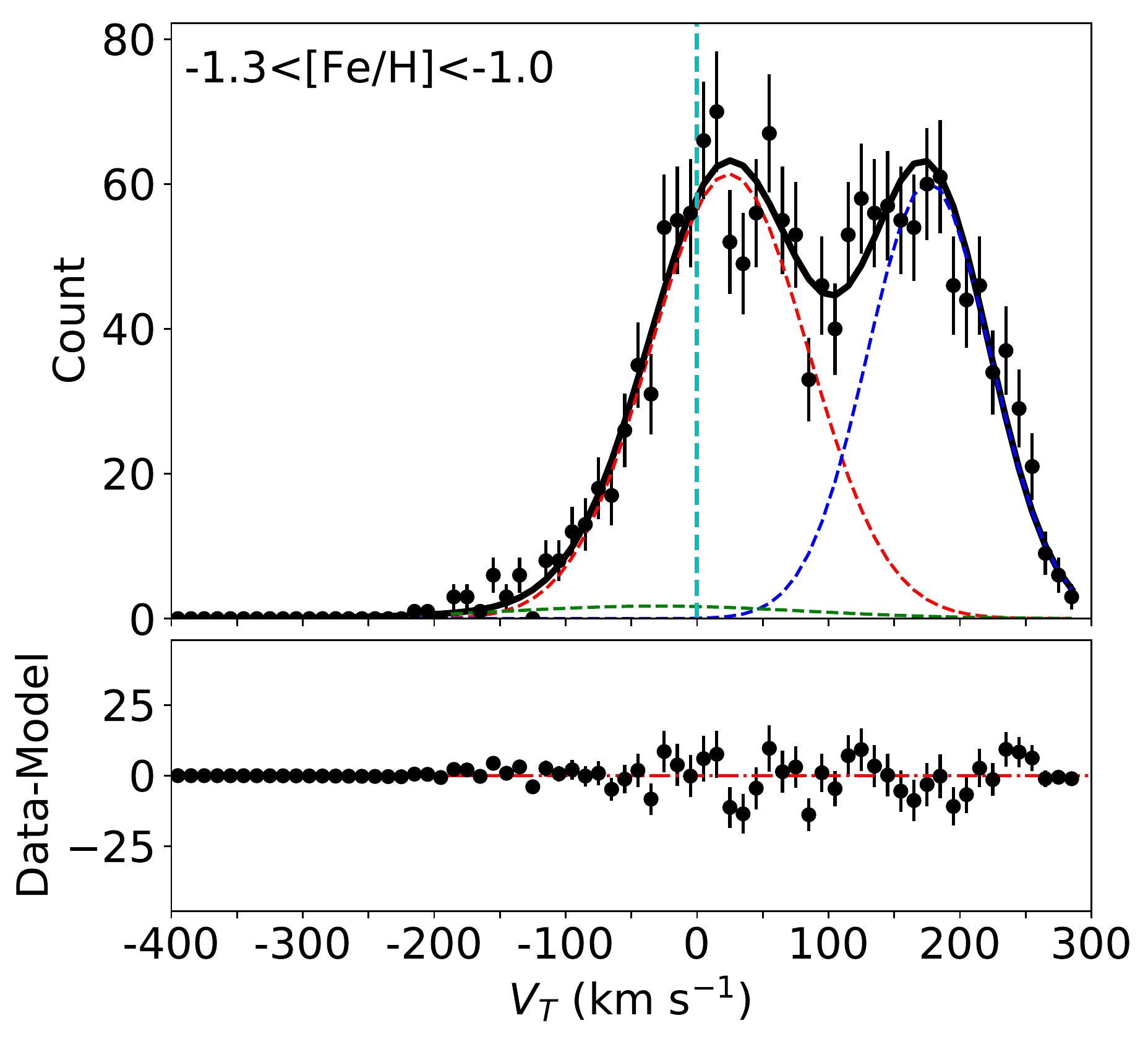}
 \end{minipage}
 \caption{ The top-left panel shows the distribution of $V_T$ for all the sample stars (black dots).
  The error bars are derived from Poisson distribution. The bin size is set at 10\,km\,s$^{-1}$. 
  The red, blue, and green dashed lines indicate the best-fit Gaussian components 
  $i=1$, $2$, and $3$ respectively. The thick black solid line indicates the total distribution
   of the three-Gaussian model. The lower plot in the panel shows the residual distribution between 
   the data and the model. No significant substructure is found in the residuals. The other 
   panels are similar to the top-left one but for different [Fe/H] bins. From  top-right to the 
   bottom-right, the panels show the distribution of $V_T$ and the corresponding best-fit 
   models with metallicity bins of -2.5$<$[Fe/H]$<$-1.6,
    -1.6$<$[Fe/H]$<$-1.3, and -1.3$<$[Fe/H]$<$-1.0, respectively.}
 \label{fig:VT_FEH}
\end{figure*}

In Figure~\ref{fig:VT_FEH}, it is seen that the distribution of $V_T$ for the samples 
(black dots with error bars) shows two peaks and a long tail beyond $V_T<-200$\,km\,s$^{-1}$. 
This hints that it may contain three components, The most dominated component shows up a peak 
located just at the right side of $V_T=0$. The second component contributes a metal-rich peak at 
around 200\,km\,s$^{-1}$.
 And the third one, maybe a broad one, contributes to the long tail at
 $V_T<-200$\,km\,s$^{-1}$, especially at metal poor end. By convenience, we assume all the three components are Gaussians
  and their distribution of $V_T$ can be defined as
\begin{equation}\label{equ:gauss}
\begin{split}
p_i(V_T^{(k)}|&f_i,V_{T,i},\sigma_i,\epsilon^{(k)})=\\
&{\frac{f_i}{\sqrt{2\pi(\sigma_i^2+\epsilon_k^2)}}}\exp\left(-\frac{(V_T^k-V_{T,i})^2}{2(\sigma_i^2+\epsilon_k^2)}\right),	
\end{split}
\end{equation}
where $i=1,2,3$ represent for the three components, $f_i$, $V_{T,i}$, and $\sigma_i$ stand for
 the fraction, mean velocity, and velocity dispersion for the $i$th component, respectively.  $\epsilon_k$ is  the 
 error of $V_T$ for $k$th star.
According to Bayes' theorem, the posterior probability distribution of the unknown parameters can be written as
\begin{equation}\label{equ:bayes}
	p(\theta_1, \theta_2, \theta_3|V_T)\propto\prod_{k=1}^{n} \sum_{i=1}^3p_i(V_T|\theta_i,\epsilon^{(k)}) p(\theta_1,\theta_2,\theta_3),
\end{equation}
where $\theta_i=(f_i,V_{T,i},\sigma_i)$ is the paramter vector for the $i$th component.
 $k$ represents for the $k$th star in the samples with totally
 $n$ stars. We adopt simplistic and loose  uniformly distributed
 priors such that $0<f_i<1$, $-100<V_{T,1}<100$\,km\,s$^{-1}$, 
 $100<V_{T,2}<300$\,km\,s$^{-1}$, $-200<V_{T,3}<0$\,km\,s$^{-1}$, and $\sigma_i<200$\,km\,s$^{-1}$. 
 We further requires $f_1+f_2+f_3=1$ so that $f_3$ is not a free parameter but derived from $f_1$ 
 and $f_2$. Therefore, we totally have 8 free parameters in the model, i.e. $f_1$, $f_2$, $V_{T,1}$,
   $V_{T,2}$,  $V_{T,3}$, $\sigma_{T,1}$, $\sigma_{T,2}$, and $\sigma_{T,3}$. 

\begin{figure*}
\centering
 \includegraphics[width=0.95\textwidth]{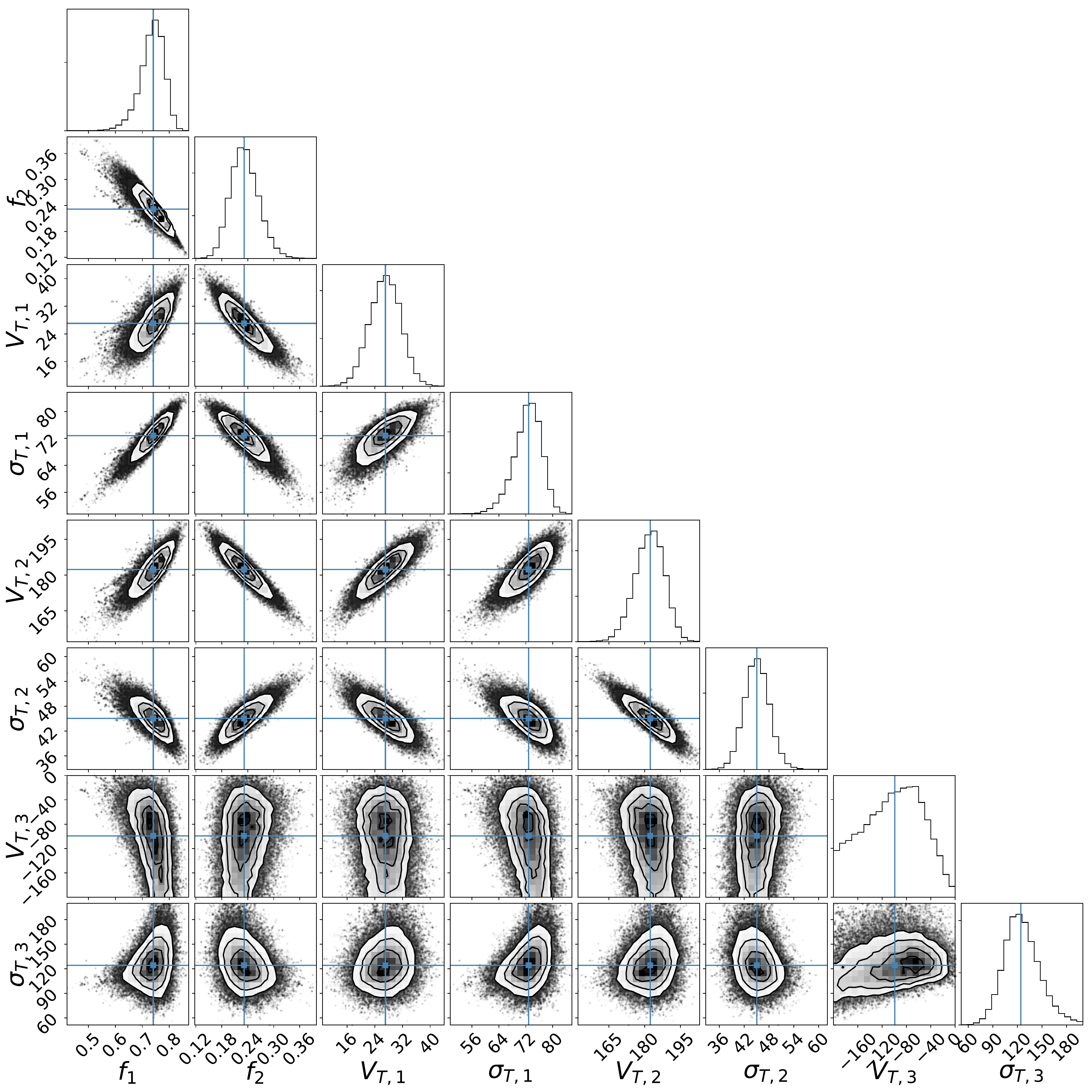}
 \caption{The results the MCMC simulation. The correlations of any two parameters are shown in the \emph{corner} plot. The cyan lines
 are the median values of each individual parameters.}
 \label{fig:results_mcmc}
\end{figure*}

Then we apply \emph{emcee} \citep{2013PASP..125..306F} to run a Markov Chain Monte Carlo 
(MCMC) simulation  with Affine solver for 
Eq. (\ref{equ:bayes})  with 2000 burn-in iterations.
 Figure~\ref{fig:results_mcmc} shows the results 
of the MCMC.  We adopt the median values of the MCMC 
samples as the best-fit parameters and 
show the best-fit models in the top-left panel of Figure~\ref{fig:VT_FEH}. 
The differences between the median values and the 
$16\%$ and $84\%$ values are calculated as the upper and lower uncertainties for each parameter.
The first row of 
Table~\ref{Tab:final} lists the best-fit parameters and their uncertainties of the three Gaussians.

\section{Result}\label{sect:result}

\subsection{Identification of the three components}
The 1st component  with $V_{T,1}=27^{+4}_{-5}$\,km\,s$^{-1}$ and 
$\sigma_{T,1}=72^{+4}_{-4}$\,km\,s$^{-1}$ is likely the stellar halo 
population, since the dispersion is quite similar to \citet{2009MNRAS.399.1223S} and
\citet{2018arXiv180504503B} at galactocentric distance
 of 10\,kpc. The 2nd component shows quite faster rotation
  with $V_{T,2}=182^{+6}_{-6}$\,km\,s$^{-1}$ and  smaller 
  dispersion $\sigma_{T,2}=45^{+3}_{-3}$\,km\,s$^{-1}$. It is noted 
  that \citet{1990AJ....100.1191M} claimed that the 
  \emph{metal-weak thick disk} has rotational velocity
  of $<V_{\phi}>=170\pm15$ km s$^{-1}$,  \citet{2014ApJ...794...58B} also obtained quite similar values
    $<V_{\phi}>=181$ km s$^{-1}$ with a dispersion 53 km s$^{-1}$ and 
    $<V_{\phi}>=166$ km s$^{-1}$ with a dispersion 47 km s$^{-1}$ using the metal weak thick disk samples 
  selected by  \citet{1973AJ.....78..687B} and \citet{2011ApJ...737....9R} respectively,
  all of which are quite consistent with our results.
   Therefore, the second component in our results is believed to be
  the metal-weak thick disk  dynamically. The 3rd 
  component shows retrograde rotation with $V_{T,3}=-99^{+47}_{-58}$\,km\,s$^{-1}$ 
  and larger dispersion of $124^{+23}_{-19}$\,km\,s$^{-1}$. As we see in 
  Section~\ref{sect:metallicity}, this population has lower 
  metallicity than the other two. It is either the so-called 
  outer halo \citep{2007Natur.450.1020C} or an accreted debris 
  in the solar neighborhood. In this work we focus on 
  the stellar halo, i.e. the 1st component.
  
\subsection{The rotation of the local stellar halo}

The stellar halo with the heliocentric distance of 4 kpc 
(the 1st component) shows a significant prograde rotation 
of 27$^{+4}_{-5}$ km s$^{-1}$. Comparing to other works, such rotation is different with the value from 
 \citet{2017MNRAS.470.1259D} and the value for K giant samples from \citet{2017MNRAS.470.2959K}.
  What should be kept in mind is that their  samples 
   have no overlapping with ours.
  \citet{1990AJ....100.1191M} 
 also studied the halo and disk with K giant stars within 4 kpc and found 
 that the stellar halo with [Fe/H]$<-1.6$ progradely rotates 
 with 25$\pm$15 km s$^{-1}$, which is quite consistent with ours,
  while the dispersion they detected 
 is at 98$\pm$13 km s$^{-1}$, slightly larger than this work.
   Considering the recent work by \citet{2018MNRAS.478..611B},
 where the volumes selected overlap with ours, the rotational velocity are quite similar with ours, 
 20 to 30 km s$^{-1}$ in spherical coordinate.
 

\begin{table*}
\begin{center}
\caption{Best fit parameters of 3-Gaussian Models derived from MCMC.\label{Tab:final}}
\setlength{\tabcolsep}{3pt}
\small
 \begin{tabular}{c|c|c|c|c|c|c|c|c|c|c}
  \hline
                    &$f_1$                       & $f_2$                         &$f_3$ &$V_{T,1}$      &$\sigma_{T,1}$    &$V_{T,2}$        &$\sigma_{T,2}$   &$V_{T,3}$         &$\sigma_{T,3}$    & $N$ \\
                    &                            &                               &      &km s$^{-1}$    &km s$^{-1}$       &km s$^{-1}$      &km s$^{-1}$      &km s$^{-1}$       &km s$^{-1}$       & \\  \hline
all                 &0.741$^{+0.038}_{-0.047}$   &    0.231$^{+0.034}_{-0.030}$  &0.028 &27$^{+4}_{-5}$ &72${^{+4}_{-4}}$  &182$^{+6}_{-6}$  &45$^{+3}_{-3}$   &-99$^{+47}_{-58}$ &124$^{+23}_{-19}$ & 3827\\ \hline
-2.5$<$[Fe/H]$<$-1.6&0.828$^{+0.089}_{-0.164}$   &    0.046$^{+0.041}_{-0.017}$  &0.126 &38$^{+9}_{-8}$ &85${^{+5}_{-7}}$  &212$^{+12}_{-37}$&30$^{+21}_{-9}$  &-47$^{+32}_{-58}$ &113$^{+21}_{-9}$  & 1134\\
-1.6$<$[Fe/H]$<$-1.3&0.752$^{+0.066}_{-0.071}$   &    0.217$^{+0.068}_{-0.061}$  &0.031 &11$^{+7}_{-7}$ &63${^{+5}_{-5}}$  &159$^{+18}_{-18}$&56$^{+8}_{-8}$   &-115$^{+60}_{-55}$&158$^{+25}_{-26}$ & 1020\\
-1.3$<$[Fe/H]$<$-1.0&0.550$^{+0.058}_{-0.059}$   &    0.421$^{+0.039}_{-0.043}$  &0.029 &24$^{+7}_{-6}$ &60${^{+7}_{-7}}$  &176$^{+6}_{-6}$  &47$^{+3}_{-3}$   &-31$^{+23}_{-52}$ &112$^{+25}_{-9}$  & 1673\\
 \hline 
\end{tabular}
\end{center}
\end{table*}
\subsection{$V_T$ versus metallicity}\label{sect:metallicity}
\citet{2017MNRAS.470.1259D} used accurate proper motion to study the rotational
 velocity of halo stars and argued that the prograde rotation of the metal 
 richer K giants is faster than that of the metal poorer sample. 
 \citet{2017MNRAS.470.2959K} also obtained a similar trend, but the difference 
 is not significant. In light of those  works, we detect the rotational
  velocity for the sub-samples with different metallicities.

 We separate the samples into three sub-groups with 
 $-2.5<$[Fe/H]$<-1.6$, $-1.6<$[Fe/H]$<-1.3$, and 
$-1.3<$[Fe/H]$<-1.0$. For each sub-sample, similar MCMC simulation is 
applied and the best-fit 3-Gaussian model parameters are listed in 
Table~\ref{Tab:final}. The comparison between the observed distributions
of $V_T$ and the models for different metallicities are displayed in Figure~\ref{fig:VT_FEH}.

The black line with red shadow in Figure~\ref{fig:Comparison} indicates
 $V_{T,1}$ as a function of [Fe/H] for the halo population (the 1st component). Rather than
  an increasing rotational velocity with the metallicity argued by previous
   works, this result prefers to an essentially flat relation between $V_{T,2}$ and [Fe/H]. 
   In other word, we do not find the rotation of the
    local stellar halo is substantially correlated with metallicity. 

It is worthy to note that the K giant samples used by 
\citet{2017MNRAS.470.1259D} are within 50 kpc and $|z|>4$ kpc 
and the samples used by \citet{2017MNRAS.470.2959K} are within 
17 kpc and $|z|>4$ kpc. Neither of them is spacially overlapping with our 
sample.

The fraction $f_1$ of the halo population seems also independent 
of [Fe/H], as seen in the corresponding column of Table~\ref{Tab:final}.
However, we find that $f_1$ is peaked at  -2$<$[Fe/H]$<$-1.6.
We find that 
$f_1$, the fraction 
of the halo population,  decreases with metallicity, 
because of the increasing of the disk population.
This again confirms that  the first component should be the stellar halo.
Note that the selection effect in [Fe/H] is not taken into account.

The black line in the blue shadow in Figure~\ref{fig:Comparison} shows rotational
 velocities at different metallicity bins for the disk component (the 2nd component). 
 It is seen that the uncertainties of $V_{T,2}$ become smaller at metal richer side. 
 This is because the fraction of the disk becomes larger when [Fe/H] increases, 
 as seen in the column $f_2$ of Table~\ref{Tab:final}. 
Thus more stars are involved in this component. This  supports that the 
 component has higher metallicity than the halo population.

Finally, $f_3$ column in Table~\ref{Tab:final} shows that the counter-rotational 
component occupies about  12.6\% at $-2.5<$[Fe/H]$<-1.6$ and only contributes 
2.9\% at $-1.3<$[Fe/H]$<-1.0$. This means that it is contributed by the most 
metal-poor stars. The averaged metallicity of the component may be even lower than the halo population.
 This is consistent with \cite{2007Natur.450.1020C} that the outer halo has metallicity of
  $\sim$-2.2 dex and in counter-rotation.

\subsection{Possible mechanism of the prograde rotation of the stellar halo}
The flat relationship between $V_{T,1}$ and [Fe/H] implies that the prograde 
rotation may not be originated from the sustained angular momenta of the minor
mergers, but in favor of an effect of secular evolution. Indeed, 
\citet{2013MNRAS.429.1949A} suggested that the rotating bar in the central 
region of the Galactic disk may transfer angular momenta to the halo. It is
also noted that \citet{2018MNRAS.473.1244X} found the halo is oblate with
axis-ratio lower than 0.5 at around 10 kpc, which may be related to the rotation
of the halo.  It is noted that, from Figure 2 in \citet{2018Natur.563...85H}
 the \emph {Gaia-Enceladus} contributes
a non-ignorable fraction of stars in the local volume, especially with metallicity around
-1.5, what's more, almost all of those stars are retrogradely rotating. This indicates
that the relatively slower rotation velocity in our results
 with metallicity between -1.6 and -1.3 might be caused, 
 at least partly, by the major merger event.
\subsection{Impact from parameter choices}
As the \emph{LSR} velocity brings a directly change in the rotational
 velocity during the conversion from proper motions, radial velocities
and distances, we test the effect of different choices of $V_{LSR}$. 
We select three $V_{LSR}$ values from literatures, 
240 km s$^{-1}$ \citep{2014ApJ...783..130R}, 232 km s$^{-1}$ \citep{2010MNRAS.403.1829S}
and 220 km s$^{-1}$ \citep{2000AJ....119..800D} for the test.
We obtain that $V_T=35^{+4}_{-5}$, $27^{+4}_{-5}$, and $15^{+4}_{-5}$\,km\,s$^{-1}$ for 
$V_{LSR}=240$, $232$, and $220$\,km\,s$^{-1}$, respectively.
It is clear that a larger $V_{LSR}$ leads to a higher halo rotation speed. However, even with the 
minimum value of $V_{LSR}$=220 km s$^{-1}$, the stellar halo is still in 
substantially prograde rotation.
Moreover, we also find that the
rotational velocity is essentially independent of the metallicity with different $V_{LSR}$.

\section{Conclusion}
\label{sect:conclusion}
We have studied the rotation of the halo with local K giant star samples within 4 kpc.
 Using the proper motions and radial velocities from \emph{Gaia} and metallicities 
 from LAMOST, we draw the following conclusions.

Firstly, the rotational velocity of the local stellar halo is strongly 
correlated to the \emph{LSR} speed, $V_{LSR}$, and also the azimuthal velocity of the 
Sun, $V_{\odot}$. With $V_{LSR}=232$  km s$^{-1}$ and $V_{\odot}$=12.24 km s$^{-1}$, 
the halo is progradely rotating with $V_T=27^{+4}_{-5}$ km s$^{-1}$.  The dispersion of
 $V_T$ of the halo K giant stars is $\sigma_{V_T}=72^{+4}_{-4}$ km s$^{-1}$. 

Secondly, we find that the rotational velocity of the stellar halo is independent of
 the metallicity in the local volume, which is different with the results in the 
 outer volume claimed by \citet{2017MNRAS.470.1259D} and \citet{2017MNRAS.470.2959K}.
 The flat relationship between $V_T$ and [Fe/H] hints that the rotation may be due to 
 secular rotation, rather than due to the net angular momenta from the minor mergers.

Finally, we also identified a metal-poor and counter-rotating hot component with 
rotational velocity of $-99^{+47}_{-58}$\,km\,s$^{-1}$, which is likely the outer halo.
And a disk-like component rotating with $V_T$=182$^{+6}_{-6}$ km s$^{-1}$ is also identified,
which is likely the metal-weak thick disk  by comparing the results with previous results
 \citep{1990AJ....100.1191M, 2014ApJ...794...58B}.

\acknowledgments
This work is supported by the National Key Basic Research Program of China 2014CB845700. 
CL acknowledges the NSFC under grants 11873057 and 11333003. 
 The LAMOST FELLOWSHIP is supported by Special Funding
 for Advanced Users, budgeted and administrated by Center for Astronomical Mega-Science,
 Chinese Academy of Sciences (CAMS). 
 HT is supported by the Young Researcher Grant of 
 National Astronomical Observatories, Chinese Academy of Science.
X.-X Xue thanks the "Recruitment Program
of Global Youth Experts" of China and NSFC 11390371.
Guoshoujing Telescope (the Large Sky Area Multi-Object Fiber Spectroscopic Telescope 
LAMOST) is a National Major Scientific Project built by the Chinese Academy of Sciences.
 Funding for the project has been provided by the National Development and Reform 
 Commission. LAMOST is operated and managed by the National Astronomical Observatories, 
 Chinese Academy of Sciences.
This work has made use of data from the European Space Agency (ESA)
mission {\it Gaia} (\url{https://www.cosmos.esa.int/gaia}), processed by
the {\it Gaia} Data Processing and Analysis Consortium (DPAC,
\url{https://www.cosmos.esa.int/web/gaia/dpac/consortium}). Funding
for the DPAC has been provided by national institutions, in particular
the institutions participating in the {\it Gaia} Multilateral Agreement.


\end{document}